\documentclass{article}
\AtBeginDocument{%
  \providecommand\BibTeX{{%
    \normalfont B\kern-0.5em{\scshape i\kern-0.25em b}\kern-0.8em\TeX}}}

\usepackage{PRIMEarxiv}
\usepackage[utf8]{inputenc}
\usepackage{xcolor}
\usepackage{amsmath}
\usepackage{amssymb,amsfonts}
\usepackage{hyperref}
\usepackage{color}
\usepackage{listings}
\usepackage{acronym}
\usepackage[T1]{fontenc}
\usepackage{lmodern}
\usepackage{multirow}
\usepackage{colortbl}
\usepackage{url}
\usepackage[export]{adjustbox}
\usepackage{colortbl}
\usepackage{tcolorbox}
\usepackage{lipsum}
\usepackage{multicol}
\usepackage{float}
\usepackage{subcaption}
\usepackage{enumerate}

\definecolor{codegreen}{rgb}{0,0.6,0}
\definecolor{codegray}{rgb}{0.5,0.5,0.5}
\definecolor{codepurple}{rgb}{0.58,0,0.82}
\definecolor{backcolour}{rgb}{0.95,0.95,0.92}
\definecolor{lightgray}{rgb}{.9,.9,.9}
\definecolor{darkgray}{rgb}{.4,.4,.4}
\definecolor{darkgreen}{rgb}{0, 0.39, 0.00}
\definecolor{Gray}{gray}{0.7}

\lstdefinestyle{mystyle}{
    backgroundcolor=\color{backcolour},   
    commentstyle=\color{codegreen},
    keywordstyle=\color{magenta},
    numberstyle=\tiny\color{codegray},
    stringstyle=\color{codepurple},
    basicstyle=\ttfamily\footnotesize,
    breakatwhitespace=false,         
    breaklines=true,                 
    captionpos=b,                    
    keepspaces=true,                 
    numbers=left,                    
    numbersep=5pt,                  
    showspaces=false,                
    showstringspaces=false,
    showtabs=false,                  
    tabsize=2https://www.overleaf.com/project
}

\acrodef{ECU}{Electronic Control Unit}
\acrodef{CACC}{Cooperative Adaptive Cruise Control}
\acrodef{RSU}{Road Side Unit}
\acrodef{SC}{Service Center}
\acrodef{ITS}{Intelligent Transportation System}
\acrodef{OBU}{On-Board Unit}
\acrodef{OBD}{On-Board Diagnostics}
\acrodef{CAN}{Controller Area Network}
\acrodef{DoS}{Denial of Service}
\acrodef{AAA}{Automobile Association of America}
\acrodef{V2V}{Vehicle-to-Vehicle}
\acrodef{V2I}{Vehicle-to-Infrastructure}
\acrodef{PID}{Priority ID}
\acrodef{CIDS}{Clock-based IDS}
\acrodef{RLS}{Recursive Least Squares}
\acrodef{IDS}{Intrusion Detection System}
\acrodef{IVI}{In Vehicle Infotainment}
\acrodef{HMM}{Hidden Markov Model}
\acrodef{ML}{Machine Learning}
\acrodef{SAE}{Society of Automotive Engineers}
\acrodef{RPM}{Revolutions Per Minute}
\acrodef{CSMA/CD}{Carrier Sense Multiple Access with Collision Detection} 
\acrodef{ML}{Machine Learning}
\acrodef{NN}{Neural Networks}
\acrodef{TPM}{Tire Pressure Monitoring System}
\acrodef{GPS}{Global Positioning System}
\acrodef{DNN}{Deep Neural Network}
\acrodef{LSTM}{Long Short-Term Memory}
\acrodef{MSG}{Messages-Sequence Graph}
\acrodef{CPD}{Change Point Detection}
\acrodef{RNN}{Recurrent Neural Networks}
\acrodef{GID}{Generative Adversarial Nets based Intrusion  Detection System}
\acrodef{CNN}{Convolution Neural Networks}
\acrodef{GRU}{Gated Recurrent Unit}
\acrodef{GAN}{Generative Adversarial Nets}
\acrodef{OBD}{On-Board Diagnostics}
\acrodef{RT}{Real-Time}
\acrodef{ACC}{Adaptive Cruise Control }
\acrodef{DDoS}{Distribute Denial of Service}
\acrodef{CPS}{Cyber Physical System}
\acrodef{IDS}{Intrusion Detection System}
\acrodef{CAN}{Controller Area Network}
\acrodef{HMM}{Hodden Markov Model}
\acrodef{SAE}{Society of Automotive Engineers}
\acrodef{IPC}{Inter-Process Communication}
\acrodef{PID} {Proportional-Integral-Derivative}
\acrodef{ADAS}{Advanced Driver-Assistance System}
\acrodef{MPC}{Model Predictive Controller}
\acrodef{DOS}{Denial of Service}
\acrodef{IQR}{linear-quadratic optimal control}
\acrodef{IVN}{In-vehicle Network}
\acrodef{SSD}{Safe Stopping Distance}
\acrodef{EKF}{Extended Kalman Filter}

\title{Improvement and Evaluation of Resilience of Adaptive Cruise Control Against Spoofing Attacks Using Intrusion Detection System}

\author{Mubark B Jedh\\
  Iowa State University\\
  Ames, United States\\
\texttt{mjedh@iastate.edu}\\
    \And
Lotfi ben Othmane\\
University of North Texas\\
Denton, United States\\
\texttt{lotfi.benothmane@unt.edu}\\
    \And
Arun K Somani\\
Iowa State University\\
Ames, United State\\
\texttt{run@iastate.edu}
}

\begin{document}

\maketitle

\begin{abstract}
The \acf{ACC} system automatically adjusts the vehicle speed to maintain a safe distance between the vehicle and the lead (ahead) vehicle. The controller's decision to accelerate or decelerate is computed using the target speed of the vehicle and the difference between the vehicle's distance to the lead vehicle and the safe distance from that vehicle. Spoofing the vehicle speed communicated through the \ac{CAN} of the vehicle impacts negatively the capability of the \ac{ACC} (Proportional-Integral-Derivative variant) to prevent crashes with the lead vehicle. The paper reports about extending the \ac{ACC} with a real-time \ac{IDS} capable of detecting speed spoofing attacks with reasonable response time and detection rate, and simulating the proposed extension using the CARLA simulation platform. The results of the simulation are: (1) spoofing the vehicle speed can foil the \ac{ACC} to falsely accelerate, causing accidents, and (2) extending \ac{ACC} with ML-based \ac{IDS} to trigger the brakes when an accident is imminent may mitigate the problem. The findings suggest exploring the capabilities of ML-based \ac{IDS} to support the resilience mechanisms in mitigating cyber-attacks on vehicles. 
\end{abstract}

\section{Introduction}
 
\begin{figure}
    \includegraphics[width=.8\linewidth]{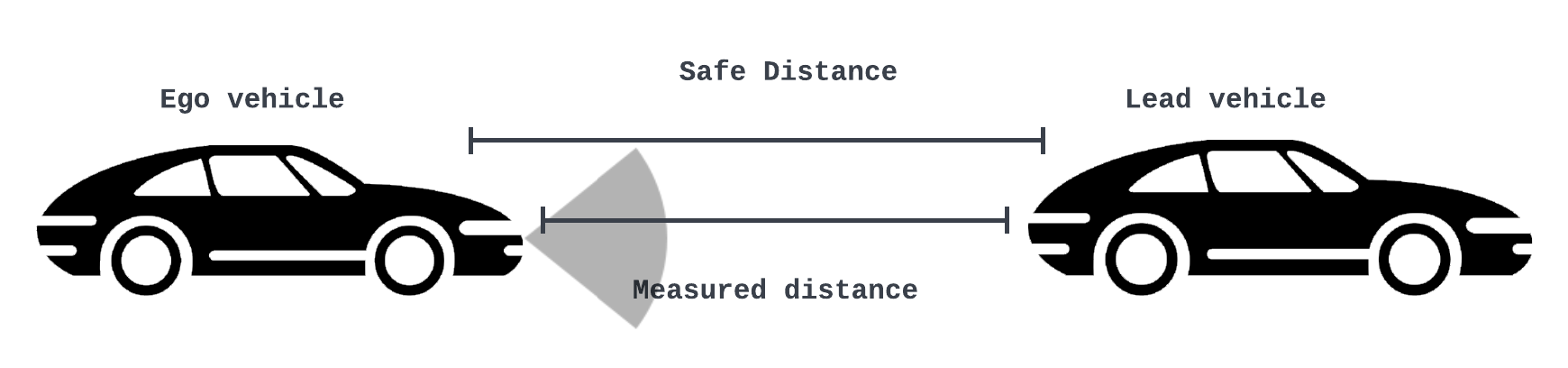}
  \caption{Visualisation of the Adaptive Cruise Control (ACC). It shows that the sensor measured distance is smaller than the safe distance.}
  \label{fig:ACCoverview}
\end{figure}

The \acf{ACC} is an advanced cruise control system that automatically adjusts the vehicle speed to maintain a safe distance between the (ego) vehicle and the lead (ahead) vehicle. The objective of the \ac{ACC} system is to make the ego vehicle travels at the driver's specified speed as long as it travels at a safe distance from the lead vehicle. The \ac{ACC}-equipped vehicle uses radar sensors to measure the distance to the lead vehicle, as depicted by Figure~\ref{fig:ACCoverview}, to take proper actions (acceleration or deceleration) in order to keep a safe distance from the lead vehicle.\footnote{The lead vehicle is driving in the same lane as the ego vehicle.} Winner et al., for example, designed an \ac{ACC} control module that uses a range sensor to measure the distance between the vehicle and the lead vehicle~\cite{961010}.\footnote{\ac{CACC} uses, in addition, to the ego vehicle's speed and distance to the lead vehicle (which are used by ACC), information about the speed and location of close-by vehicles to better regulate its speed~\cite{8656970}.} The first commercial system in use was a lidar-based distance detection system Debonair Mitsubishi, which is available since 1992. 

The promises of \acf{ACC} systems in terms of driver comforts and safety assurance encouraged researchers to experiment with the feasibility and impact of cyber-attack on \ac{ACC} systems. The experiments showed that forcing the \ac{ACC} to use the wrong information about the distance between the vehicle and the lead vehicle leads to the wrong acceleration/deceleration decision, which leads to accidents. The falsification of the safe distance was achieved using two techniques: (1) forcing the distance sensor (e.g., LIDAR or front Camera) to report wrong information to the \ac{ACC}~\cite{10.1145/3319535.3339815, harris_2015}, and (2) manipulating the distance between the vehicle and the lead vehicle communicated by the distance sensor to the \ac{ACC} through the CAN Bus~\cite{9349140,7967438}. The proposed solutions focus on communicating the wrong distance between the vehicles to the \ac{ACC} by, e.g., using time-varying sampling of the distance between the two vehicles~\cite{7967438} or \ac{MPC} system~\cite{9349140}.

\begin{table*}[hbt]
\caption{Strategy for resilience.}
\label{tab:modules}
\centering
\begin{tabular}{p{0.1in} p{0.8in}  p{0.9 in} p{3.5in} p{.3in} } 
\hline
\rowcolor{Gray}\hline
ID&  Technique & Addressed security aspect& Description & Ref.\\ \hline
\rowcolor{lightgray}
1 &Recursive Least square& No Security &Online parameter estimation of distance gap by optimizing the root mean squared error (RMSE) between simulated space gap data and recorded space gap data.& \cite{9294538, 9195163}\\\hline
2 & Recursive Least square &  Challenge response authentication&A RLS is used to estimate the spoofed sensor measurement.& \cite{10.1145/3061639.3062241}\\ \hline
3 & Extended Kalman filter& Yes, through estimation&Estimating the velocity of the vehicle in the local reference system. &\cite{8804527,8814161}\\\hline
4& Kalman filter&No Security &Estimating the velocity of the vehicle in the local reference system and concurrently the absolute position & \cite{8804527} \\\hline
5 & Particle filter&No Security &Online parameter estimation of \ac{ACC} & \cite{9195163} \\\hline
6&Model Predictive Controller &Yes, through estimation & Applies the linear model of the system, disturbance, and noise models to estimate the state of the control system and also anticipate the system's future outputs. &\cite{9349140} \\\hline\hline
\end{tabular}
\end{table*}

The \ac{ACC} uses the vehicle's speed, besides the distance between the vehicle and the lead vehicle, in computing the acceleration/deceleration control decision. Forcing the \ac{ACC} to use the wrong vehicle speed could also potentially lead to the wrong acceleration/deceleration decision, which leads to accidents. This paper proposes extending the \ac{ACC} system with a real-time \ac{IDS} to detect cyber-attacks on the vehicle and to trigger the brakes when spoofing of the vehicle speed is detected. We implemented the solution into CARLA simulator~\cite{Dosovitskiy17} considering the performance of a real-time \ac{IDS} implemented in our previous research~\cite{9076852,9490207,06680}. The solution is assessed using the following scenarios: (1) simulate the ego vehicle trailing the lead vehicle and using an \ac{ACC} to avoid crashes; (2) simulate the ego vehicle trailing the lead vehicle while using the \ac{ACC} to avoid crashes, and spoofing the speed of the ego vehicle; and (3) simulate the ego vehicle trailing a lead vehicle, spoof the speed of the ego vehicle, and use an \ac{ACC} extended with a simulated real-time \ac{IDS}~\cite{9490207}.

The contributions of the paper are:
\begin{itemize}
    \item Demonstrate that spoofing the speed of the vehicle can mislead the \ac{ACC} to compute the wrong safety distance with the lead vehicles, leading to potential crashes. 
    \item Extend the {ACC} (\ac{PID} variant~\cite{rajamani2011vehicle}) with a real-time \ac{IDS} to force cold brake when spoofing of the vehicle speed is detected addresses the problem. 
\end{itemize}

The results suggest that using the vehicle's \ac{IDS} by proactively monitoring the \ac{CAN} bus will improve the resilience of the \ac{ACC} system to cyber-attacks.

The paper is organized as follows:
section~\ref{sec:relworks} gives an overview of related works; section~\ref{sec:approach} describes the proposed \ac{ACC} extension with real-time \ac{IDS}; section~\ref{sec:simulation} describes and analyses the results of simulating the proposed extension; and section~\ref{sec:Conclusions} concludes the paper.

\section{Related work}\label{sec:relworks}

This section describes related work on cyber-attacks on \ac{ACC} systems and cyber-resilience of \ac{ACC} systems. 

\subsection{Cyber-attacks on Adaptive Cruise Control}

Several researchers investigated the security of LiDAR, especially spoofing the LiDAR signal. For instance, Harris~\cite{harris_2015} developed an attack on LiDAR laser that makes the vehicle wrongly believe that there is a large object in front of it, preventing it from moving by overwhelming the LiDAR sensor~\cite{harris_2015}. Coa et al. spoofed obstacles, leading the LiDAR-based perception to believe it is close to the object~\cite{10.1145/3319535.3339815}. Also, Rad et al.~\cite{rad2020experimental} and Jagielski et al.~\cite{jagielski2018threat} showed that sensor spoofing against RADAR and LIDAR impacts the efficiency of the \ac{ACC} and \ac{CACC} leading to potential discomfort of the passenger and safety hazard including accidents. Farivar et al.~\cite{9349140} proposed a covert attack on \ac{ACC} that manipulate radar sensor input, which leads the \ac{ACC} to decrease the safe distance, causing crashes. The authors developed an \ac{IDS} for such attacks and corrected the system using \ac{MPC} system. Moreover, Sun et al. demonstrated a spoofing attack against a LiDAR sensor, effectively tricking the system into perceiving an obstacle in its path by transmitting laser signals to the victim’s LiDAR~~\cite{https://doi.org/10.48550/arxiv.2006.16974}. Their result showed that attackers can achieve 80\% mean success rate on all LiDAR target models. Petit et al.~\cite{petit2015remote} showed the efficacy of the Lidar relay attacks and spoofing attacks using a cheap transceiver. 

The proposed solutions to such attack include the physical chall-enge-response authentication (PyCRA) technique, which was developed by Shoukry et al. ~\cite{Shoukry2015PyCRAPC} to enhance the cyber-resilience of sensors to attacks. PyCRA assumes that an attacker cannot detect a challenge immediately due to its hardware and signal processing latency. Given that, PyCRA detects an attack signal that continues to be higher than a noise threshold during a challenging period using the Chi-square method. PyCRA turns off sensors that have been attacked, providing an authentication mechanism that not only detects malicious attacks but provides resilience against them.

The \ac{ACC} system uses the speed and distance sensors information to compute the desired acceleration to maintain a safe distance from the lead vehicle. The system is integrated as an embedded system into the \ac{CAN} Bus, which is known to be vulnerable to cyber-attacks exploiting the lack of secure communication between the \acp{ECU} communicating through the bus~\cite{Othmane2015}. Heijeden et al.~\cite{8275598} showed that controllers are vulnerable to jamming or \ac{DoS} and message injection attacks and proposed quantifying the impact of attacks on vehicle controllers system. Furthermore, Tianxiang et al.~\cite{7967438} studied the stability of the \ac{ACC} system subject to~\ac{DoS} by performing real-time \ac{DoS} attacks on the \ac{ACC} system at various time steps and studying the time it takes the \ac{ACC} system to return to the closed-loop system. They found that under DoS attacks, the \ac{ACC} system behaves as an open-loop system, and the speed errors increase. 

\subsection{Cyber-resilience of Adaptive Cruise Control}

 Several authors investigated different aspects of \ac{ACC} resilience to cyber-attacks. Table~\ref{tab:modules} summarizes the main proposed techniques. Oh et al. proposed using a sliding mode observer to detect sensor faults in the case of cyber-attacks on the acceleration sensor and radar~\cite{oh2018functional}. Abdollahi et al. also proposed using a sliding mode observer to detect \ac{DoS} attack and estimate correction~\cite{8293801}. Fiu et al. proposed a technique to estimate the position of a vehicle under GPS spoofing and LIDAR replay attacks~\cite{8814161}. In addition, Liu proposed the use of an extended Kalman filter to fuse sensor measurements to estimate a vehicle’s position and designed a Cumulative Sum (CUSUM) detector based on the residual of \ac{EKF} to monitor the inconsistency~\cite{8814161}. When detecting that the sensor is under cyber-attack, \ac{EKF} is reconfigured to estimate the correct position of a vehicle~\cite{8814161}.

Each of the resilience methods discussed above proposes new formulations of the controllers to mitigate specific attacks such as \ac{DOS}, which limit their adaptation in practice. Also, given that the algorithms/formulation are public, the attackers should be able to design attacks that avoid the protection--much like avoiding the failure-detection capabilities of existing controllers. This paper proposes delegating detecting cyber-attacks to \ac{IDS} and extending the \ac{ACC} controller to use \ac{IDS} to mitigate such cyber-attacks. This allows having a robust and generic solution for cyber-attacks for \ac{ACC} systems.
 
\section{Extending Adaptive Cruise Control with ML-based Intrusion Detection System}\label{sec:approach}

The negative impact of spoofing distance sensors on the efficiency of ACC started to be known, and the community started developing ACC models resilient to cyber-attacks on these sensors and the distance data communicated through the CAN bus. These models do not consider spoofing the vehicle speed, also used by the ACC system. This paper proposes a generic solution to the ineffective resilience of ACC to cyber-attacks, focusing on spoofing vehicle speed. This section describes the proposed solution.

\begin{figure}[tb]
    \centering
    \includegraphics[width=0.25\textwidth]{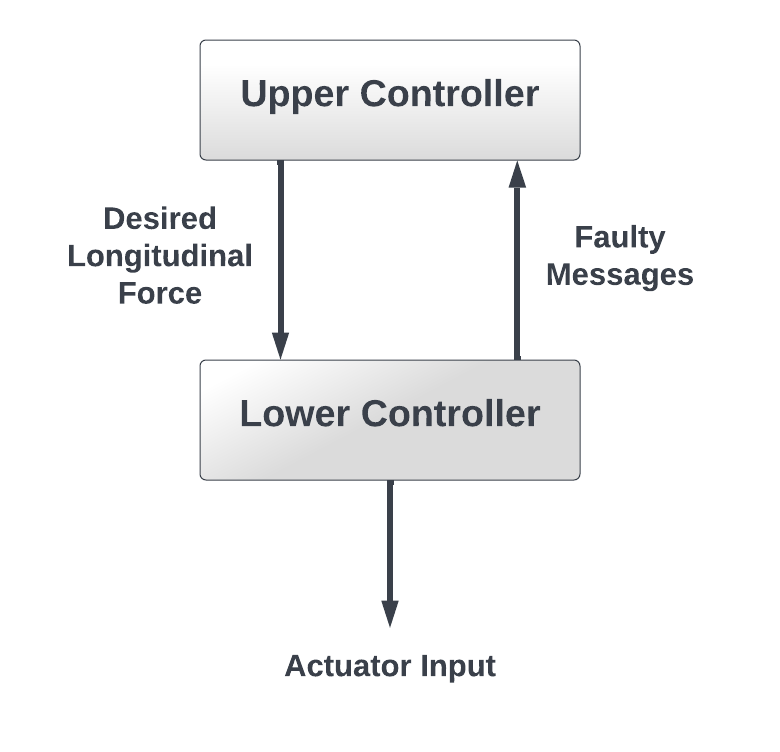}
    \caption{Controllers composing the Adaptive Cruise Control System.}
    \label{fig:lower_upper_controller}
\end{figure}

\begin{figure}[tb]
    \centering
    \includegraphics[width=0.5\textwidth]{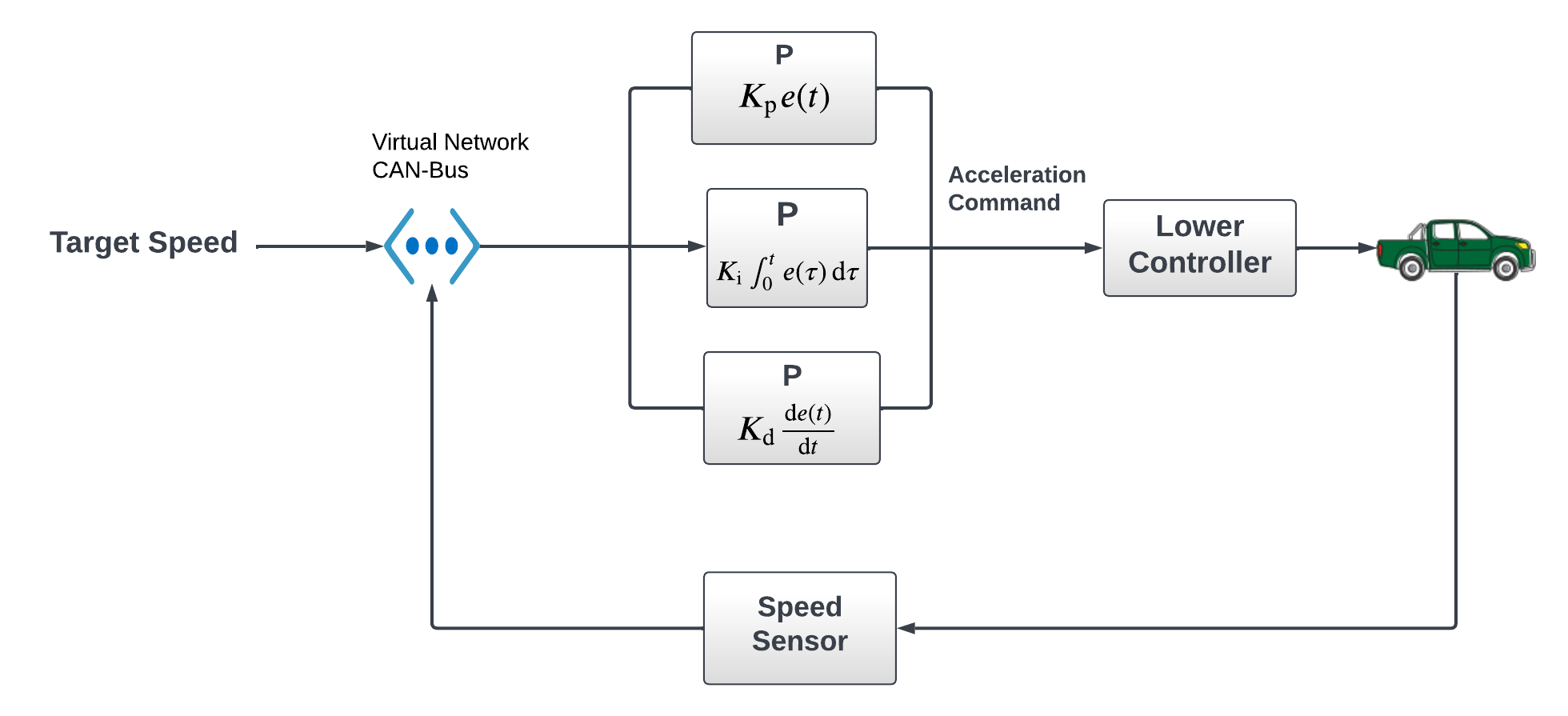}
    \caption{Components of the Proportional-Integral-Derivative (PID) Controller model.}
    \label{fig:PID}
\end{figure}

\begin{figure}[tb]
    \centering
    \includegraphics[width=0.48\textwidth]{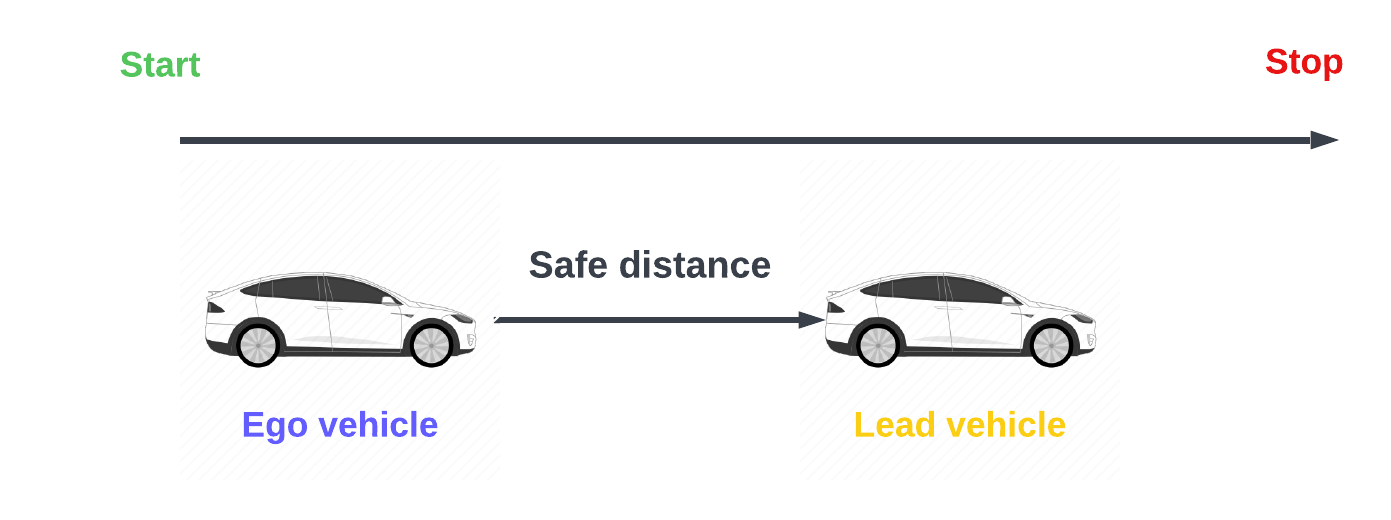}
    \caption{The Ego vehicle uses an ACC while following the lead vehicle, which allows it to keep a safe distance.}
    \label{fig:safedistance}
   
\end{figure}

\subsection{Overview of the Adaptive Cruise Control}

The primary goal of the \ac{ACC} is to maintain a safe distance from the lead vehicle, that is, keeping the difference between the current distance from the lead vehicle and the safe distance, as shown by figure~\ref{fig:safedistance}, formulated using Equation~\ref{eq:distance}, higher than zero. The \ac{SSD}, or simply the safe distance, determines how far the vehicle travels before it comes to a complete stop and avoids collision with the lead vehicle, as is formulated using Equation~\ref{eq:safe-distance}~\cite{book_policy}.

\begin{equation}
\centering
    D = D_c - SSD
    \label{eq:distance}
\end{equation}

\begin{equation}
\centering
    SSD = 0.278 \times t \times v + \frac {v^2} {254 \times {f+g}}
    \label{eq:safe-distance}
\end{equation}

where \ac{SSD} is the stopping distance in meters, $t$ is the perception-reaction time in seconds (it is 2.5s for most drivers), $v$ is the speed of the car in km/h, $G$ is the slope of the road, and $f$ is the coefficient of friction between the tires and the road. The ego vehicle avoids collision while moving by stopping immediately when the SSD is less than its current distance from the lead vehicle.

The \ac{ACC} system uses two controllers as depicted by Figure~\ref{fig:lower_upper_controller}: an upper-level controller and a lower-level controller. The lower-level controller determines the throttle and brake while the upper-level controller determines the desired longitudinal acceleration to attain the desired spacing and constant speed~\cite{rajamani2011vehicle}. Different dynamic responses that implement the two-levels \ac{ACC} system have been proposed including~\ac{PID} controllers~\cite{rajamani2011vehicle}, Linear Quadratic Regulator control (LQR)~\cite{https://doi.org/10.48550/arxiv.1911.08349}, Sliding Mode Control~\cite{6870952}, Fuzzy Logic Control~\cite{4406871}, and~\acf{MPC}~\cite{9349140}. We use in this paper the \ac{PID} model because it is widely used in many systems to reach the stability status.  

\begin{equation}
    e(t) = S_t(t) - S_c(t)\\
    \label{equa:Timeheadway}
\end{equation}

Figure~\ref{fig:PID} depicts the overall \ac{PID}-based \ac{ACC} system~\cite{rajamani2011vehicle}. The PID controller adjusts the acceleration and deceleration commands to minimize the error computed using Equation~\ref{equa:Timeheadway}, which measures the difference between the target speed of the vehicle ($S_t$) and its current speed ($S_c$) as measured by the speed sensor~\cite{rajamani2011vehicle}. The speed error is used to compute the control signal $u(t)$, shown in Figure~\ref{fig:PID}, using Equation~\ref{PID_equation}, which uses three constants: 
\begin{itemize}
    \item $K_{p}$ -- proportional gain of the action to the error,
    \item $K_{i}$ -- integral gain to reduce the steady-state errors through low-frequency compensation by an integrator,
    \item $K_{d}$ -- derivative gain to improve the transient response through high-frequency compensation by a differential.
\end{itemize}

\begin{equation}\label{PID_equation}
    u(t) = K_\text{p} e(t) + K_\text{i} \int_0^t e(t) \,\mathrm{dt} + K_\text{d} \frac{\mathrm{d}e(t)}{\mathrm{d}t} ~~\cite{rajamani2011vehicle}
\end{equation}

Increasing the $K_{p}$ value helps the vehicle reaching the target speed more quickly, but tends to exceed its target and overshoot. The $K_{d}$  term affects the decrease of the overshoot. The $K_{i}$ value affects the capability to limit the steady error and prevent oscillatory. Adjusting the gains of the $K_{p}$ , $K_{d}$, and $K_{i}$ allows to achieve a satisfactory overall response.

\subsection{Architecture of the Extended Adaptive Cruise Control}

 Resilience has been an effective active solution for vehicle controllers, providing robust capabilities for vehicles to reduce errors and detect failures during vehicle operation. Resilience mechanisms trust the sensors to provide measurements and consider the random outliers as errors. Cyber-attacks mislead the controllers by spoofing the sensor measurements. A resilient controller should enhance the system's performance during an attack. 
 
 \ac{IDS} has been proposed as a solution, although passive, to mitigate cyber-attacks on vehicles~\cite{8640808,8688625}. It detects intrusion that results in compromised system components in the \ac{ACC} system (e.g., sensors and controllers output) and reports the anomaly to combat the malicious attackers. \ac{ACC} cannot report an intrusion detected by the \ac{IDS} to a remote security expert or even the driver and wait for their decision while possibly getting closer to the lead vehicle. 

The ACC must use a correct, safe distance to work properly. We propose to extend the \ac{ACC} with a vehicle \ac{IDS} that allows detecting the vehicle's speed spoofing with reasonable efficacy. The \ac{ACC} has two options to mitigate detected cyber-attacks: (1) trigger cold break and (2) get the correct speed using another approach, including predicting it using machine learning models. We use option one in this paper as it allows us to evaluate the solution more easily.      

\begin{figure}[tb]
    \centering
    \includegraphics[width=0.50\textwidth]{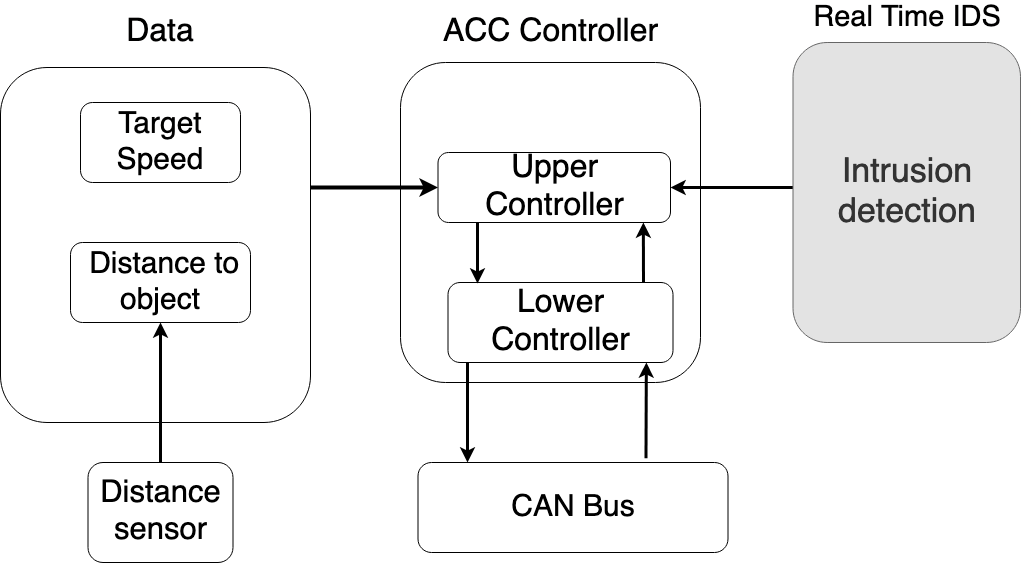}
    \caption{Architecture design simulation setup for scenario three} 
    \label{fig:scenario_three_arachetiture}
\end{figure}

\begin{figure}[tb]
    \centering
    \includegraphics[width=0.50\textwidth]{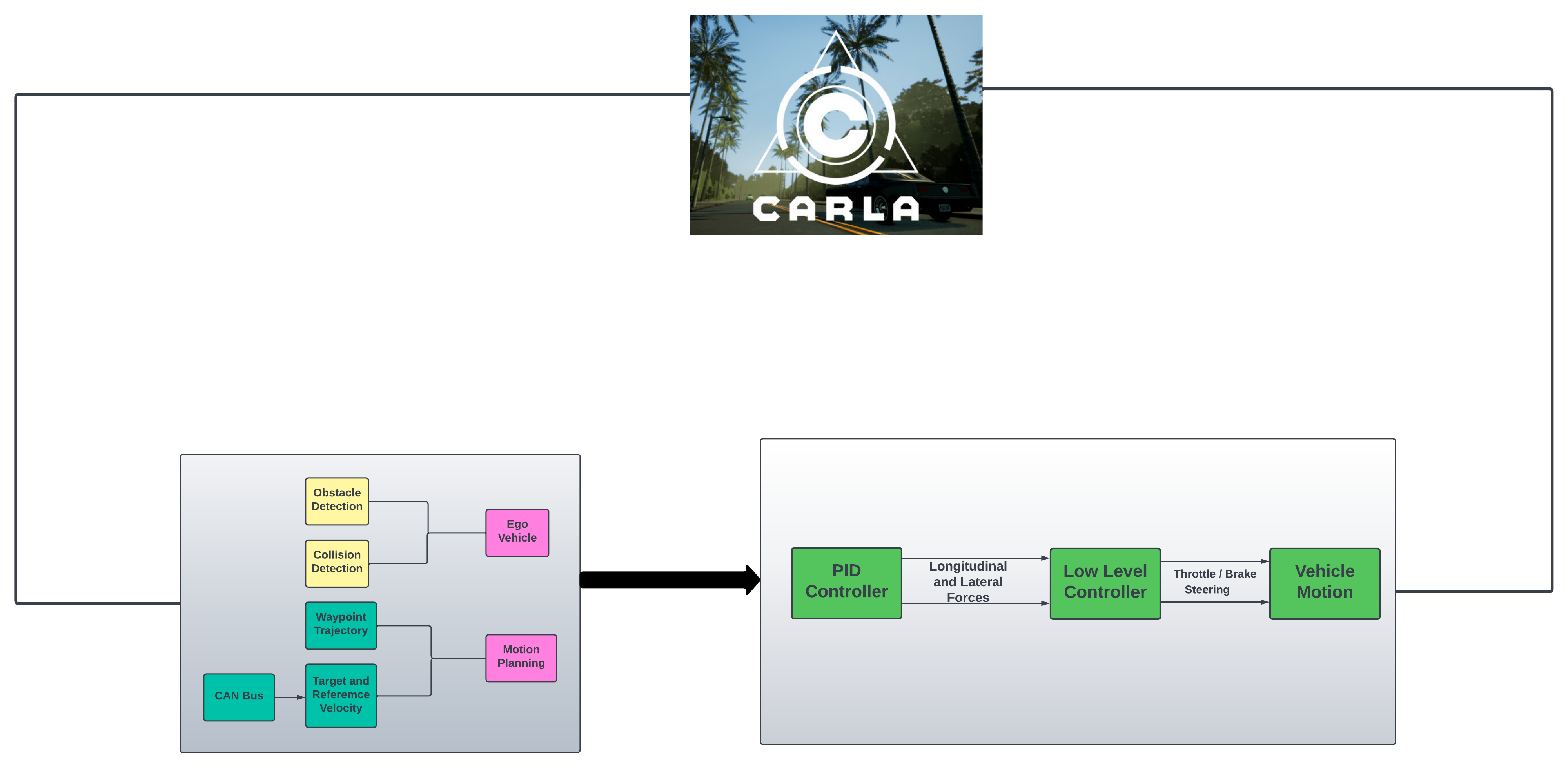}
    \caption{Architecture for simulation setup } 
    \label{fig:arach}
   
\end{figure}

Figure~\ref{fig:arach} shows the architecture of the proposed extended \ac{ACC}, called from now on ACC-IDS, that uses a real-time IDS to detect attacks~\cite{06680}, in addition to the driver-specified target speed, the distance to the lead vehicle, and the vehicle speed (used to compute the safe distance).\footnote{Let's assume that spoofing the road's slope and the friction coefficient between the tires and the road may not have an important impact on the safe stopping distance.} Real-time \ac{IDS} for cars monitors the \ac{CAN} bus of the vehicle and detects message injections that spoof, for example, the speed sensor using, e.g., machine learning techniques.   

\begin{figure}[tb]
    \centering
    \includegraphics[width=0.50\textwidth]{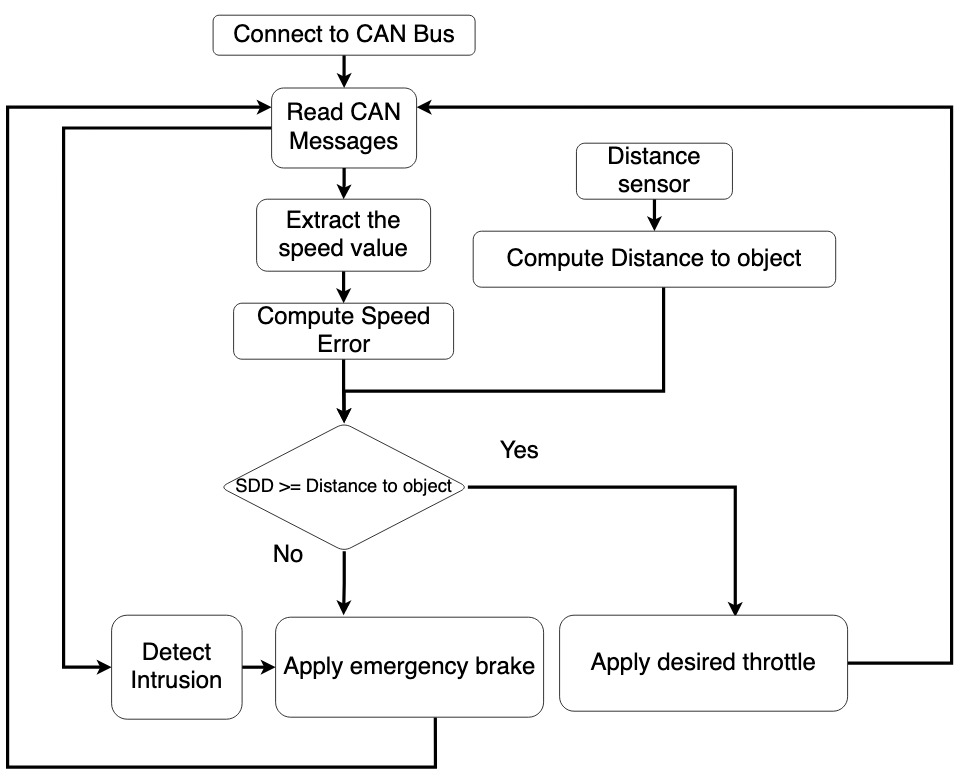}
    \caption{Flowchart of the proposed extended ACC, called ACC-IDS. The IDS components are marked with the gray color. } 
    \label{fig:flowchatextendedACC}
\end{figure}

\begin{table}[tbp]
\caption{The performance of the \ac{IDS} in the simulated scenario~\cite{06680}.}
\label{tab:simulationvsonline}
\centering
\begin{tabular}{p{2.0in}p{2.0in}}
\rowcolor{Gray}\hline
  Factor  & Parameter\\\hline
Dataset  &  Simulated real-time \\ \hline
Network 	& Simulated \ac{CAN} bus\\ \hline
Detection rate		&0.97\%\\ \hline
Detection latency  &152 ms\\ \hline
Response time 	& 1026 ms  \\ \hline\hline

\end{tabular}
\vspace{-0.20 in}
\end{table}

Figure~\ref{fig:flowchatextendedACC} shows the flowchart of the proposed \ac{ACC}-\ac{IDS}. It shows that the messages read from the \ac{CAN} bus of the vehicle are processed by the \ac{IDS} component and also used to check whether the vehicle maintains a safe distance from the lead vehicle. The ACC-IDS applies an emergency brake when either the distance to the object is less than the SSD or an intrusion is detected. 

The ACC-IDS can be effective only if the IDS provides high accuracy in detecting messages injections\footnote{Messages are injected as spoofing of a given ECU.} and 
 low response time sufficient for the upper and lower controllers to adjust the vehicle's speed and avoid an accident. There is currently no real-time IDS for connected vehicles to use in evaluating the solution. To assess the solution, we use the performance measures of a real-time \ac{IDS} that we have proposed and simulated~\cite{9490207,06680} using data collected from a moving vehicle under malicious speed reading message injections~\cite{9076852}. Table~\ref{tab:simulationvsonline} provides the performance measurement of the real-time IDS deployed on a Raspberry Pi in a simulated environment~\cite{06680}. Notably, the response time of the IDS is 1.026 seconds, which is lower than the driver's reaction time of 2.5 seconds.

\section{Evaluation of the ACC-IDS using CARLA Simulation}\label{sec:simulation}

We evaluate the ACC-IDS solution using the simulation method and CARLA simulation environment tool.

\subsection{Simulation Setup}

In the simulation, the position of the ego vehicle and lead vehicle is set in defined coordinates at the start point with a safe distance of 30 m. The ego vehicle detects the positions of neighboring vehicles and obstacles in real time.\footnote{The vehicle is allowed to ignore the speed limits and traffic lights. The stop destination is straight ahead of the starting point, and there are no dynamic objects in the environment.} The \ac{PID} controllers compute both the steering and acceleration/braking commands in order to achieve a safe distance from the lead vehicle.\footnote{It is assumed that the lower-level controller applies controller output synchronously, which makes the vehicle accelerate or decelerate exactly with the desired acceleration.} The ego vehicle uses obstacle detection and safe distance to avoid crashes. The simulated scenarios are:

\begin{enumerate}
\item Scenario 1 - Ego vehicle drives at target speed of 25 km/h and lead vehicle drives at speed of 0 km/h to 30 km/h and back to 0 km/h.\footnote{The speed of the lead vehicle is read from a CAN bus log dataset of a moving vehicle~\cite{9076852}.} The ego vehicle sensor speed is not spoofed
\item Scenario 2 - The target speed of the ego vehicle and lead vehicle are set to 60 km/h. The ego vehicle sensor speed is spoofed to 10km/h. The \ac{ACC} of the ego vehicle does not use an \ac{IDS} to detect the spoofing.  The attack interval is fixed at a rate of 75\%, leading the controller to falsely produce more throttle force than sufficient. The scenario is repeated with the target speed of the ego vehicle and lead vehicle is set to 90 km/h.
\item Scenario 3 - The target speed of the ego vehicle and lead vehicle is set to 60 km/h. The ego vehicle sensor speed is spoofed to 10km/h. The \ac{ACC} of the ego vehicle uses the real-time \ac{IDS} to detect the spoofing. The parameters of the \ac{IDS} are provided by Table~\ref{tab:simulationvsonline}. The scenario is repeated with the target speed of the ego vehicle and lead vehicle is set to 90 km/h.
\end{enumerate}

We note that we also simulated the three scenarios with the target speed of the ego vehicle and lead vehicle is set to 40 km/h~\cite{mubarek2023}. The results are omitted as it provides less insight as setting target speed of 60 km/h and 90 km/h.

We designed our simulation scenario using the CARLA simulator~\cite{Dosovitskiy17}. The proposed evaluation framework, including the structure of the simulation, is shown in Figure~\ref{fig:arach}. Data from the sensors are obtained by connecting to the CARLA server. We used {\tt Town04} and {\tt Town05} given by the CARLA simulator to model both vehicles' routes. CARLA uses a set of \acp{ECU} emulators. Our simulation sends CARLA sensor data as messages periodically into a Linux virtual CAN bus. It also injects "spoofed" messages into the virtual CAN Bus to mislead the moving vehicle. CARLA's ACC and the simulated IDS read the \acp{ECU} data from the virtual CAN bus and process them according to the flowchart of Figure~\ref{fig:flowchatextendedACC}. 

We used in this simulation the controller' constants used by the default CARLA setup. Given that driving is in a plane, the upper controller uses two controllers: (1) a longitudinal controller and (2) a latitude controller. We used  $K_{p}$ = 1.0, $K_{d}$ = 0, $K_{i}$ = 0.7, and $dt$ = 0.05 for the longitudinal controller and $K_{p}$ = 1.98, $K_{d}$ = 0.20, $K_{i}$ = 0.07, and $dt$ = 0.05.\- 
for the lateral controller.

We implemented and deployed the simulation in accelerated computing AWS instances with 12 core processors, 200 GB of memory, and 8 GB of dedicated Nivida GPU running with a 64-bit Ubuntu 18.04.3 LTS. 

\subsection{Result and Analysis of the Simulation Experiments}\label{sec:result}

This subsection reports about the results of the simulation scenarios.

\begin{table}[tbp]
\caption{Summary of the simulated scenarios. The ego vehicle uses \ac{PID}-based \ac{ACC} for all the scenarios.}
\label{tab:simulationresults}
\centering
\begin{tabular}{p{0.4in} p{1.1in}p{1.3in} p{1.5in}p{0.7in}p{0.6in}}
\rowcolor{Gray}\hline
   Scena-rio& Target speed of ego vehicle&Target speed of lead& Attack type &Use of IDS & Crashes \\ \hline
1 &  25 km/h & extracted from the CAN Bus& no spoofing& No& No  \\\hline
2 & 60 and 90 km/h & 60, and 90 km/h& spoofing of speed sensor & No & yes \\\hline
3 & 60 and 90 km/h & 60,and 90 km/h& spoofing of speed sensor & yes & No \\\hline
\hline
\end{tabular}
\vspace{-0.1in}
\end{table}

\begin{figure*}[tbp]
    \centering
     \begin{subfigure}[t]{0.33\textwidth}
        \centering
     \includegraphics[width=1.0\linewidth]{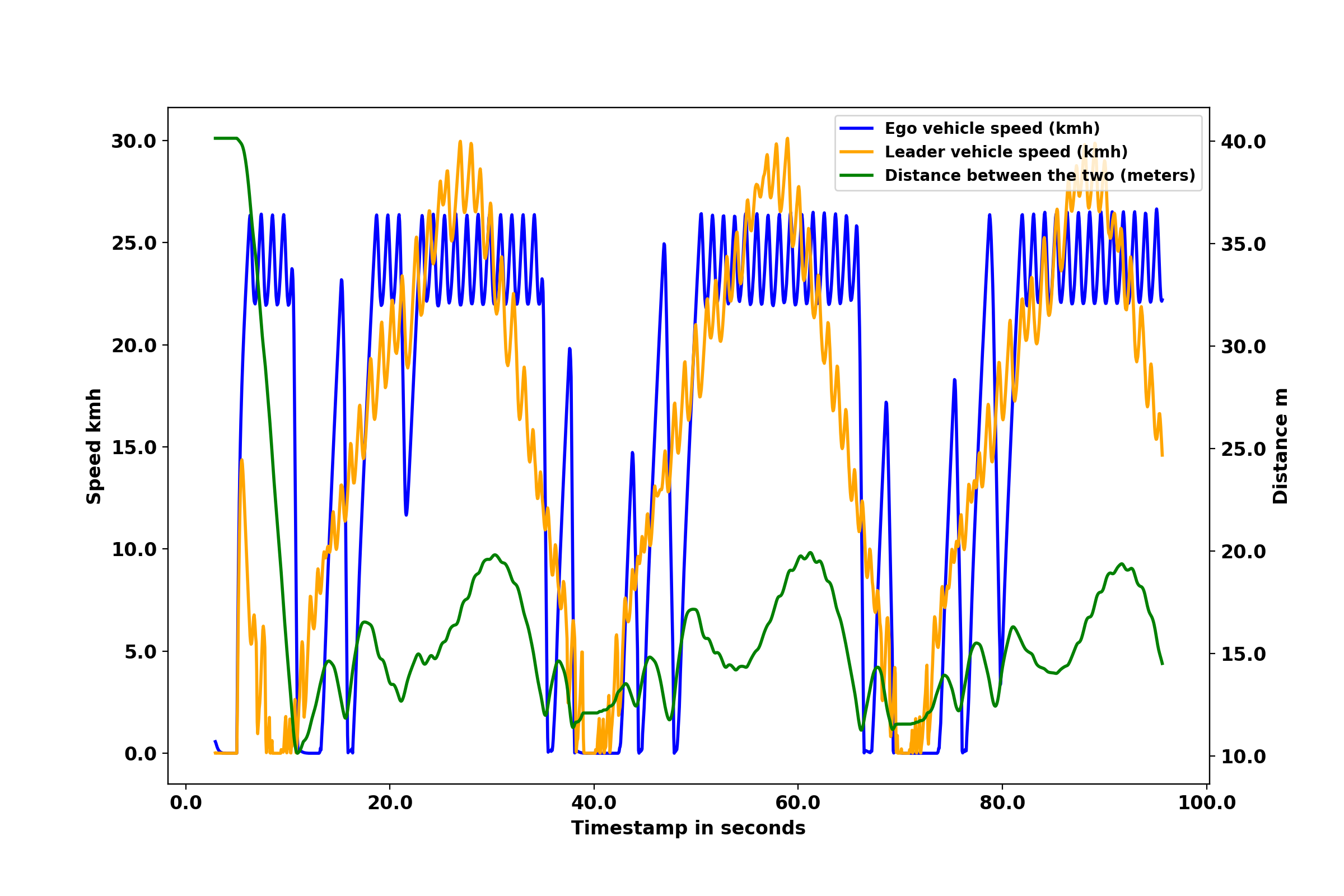}
    \centering
    \caption{The ego vehicle mimics a car driving at speed increasing from 0 km/h to 30 km/h and decreasing to 0km/h again. The ECUs of the vehicle are not spoofed.}
    \label{fig:1st_normal}
    \end{subfigure}%
    ~~
     \begin{subfigure}[t]{0.33\textwidth}
        \centering
     \includegraphics[width=1.0\linewidth]{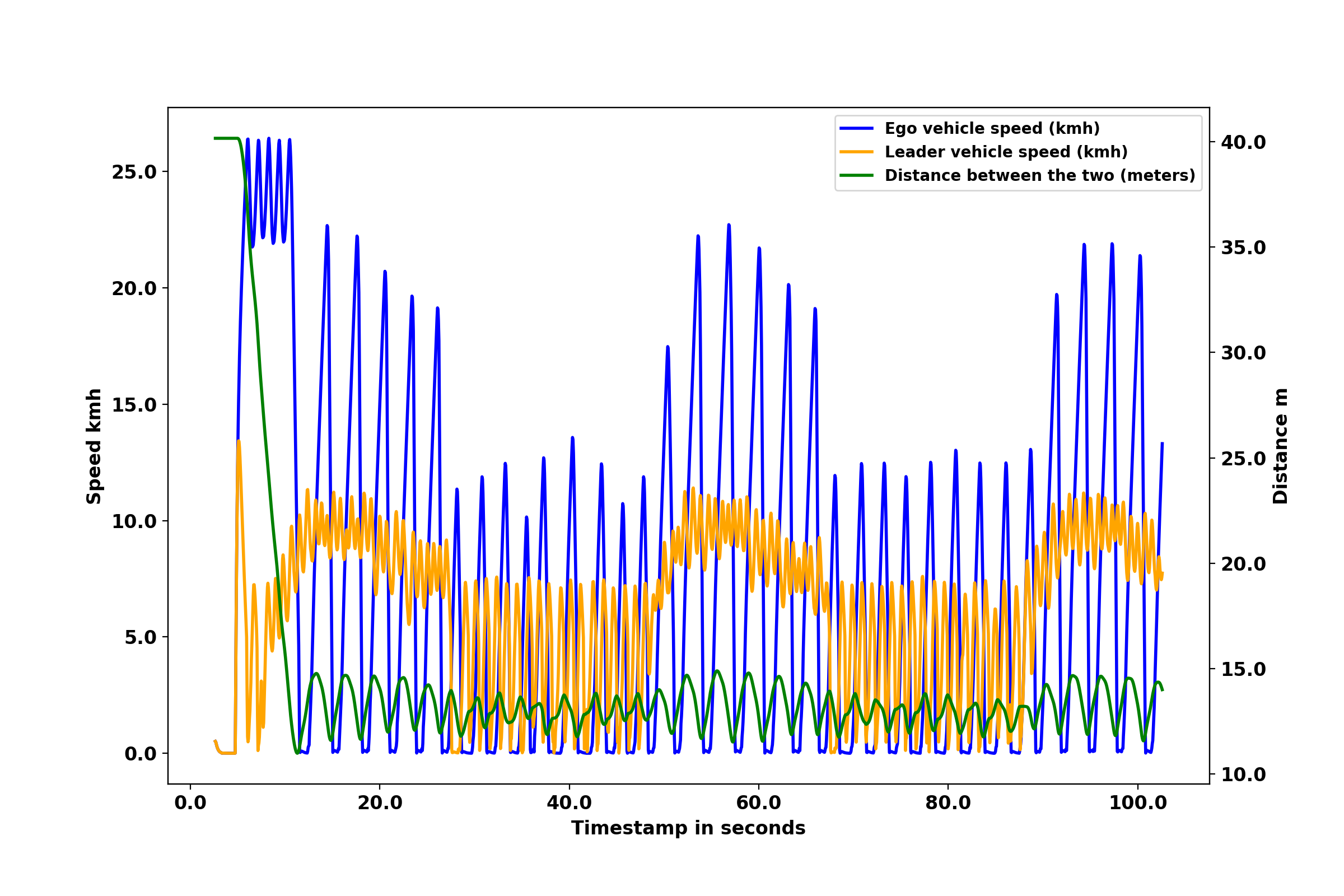}
    \centering
    \caption{The ego vehicle uses the CAN bus log of moving vehicle while the RPM reading is being spoofed. }
    \label{fig:1st_rpm}
    \end{subfigure}%
      ~~
     \begin{subfigure}[t]{0.33\textwidth}
        \centering
     \includegraphics[width=1.0\linewidth]{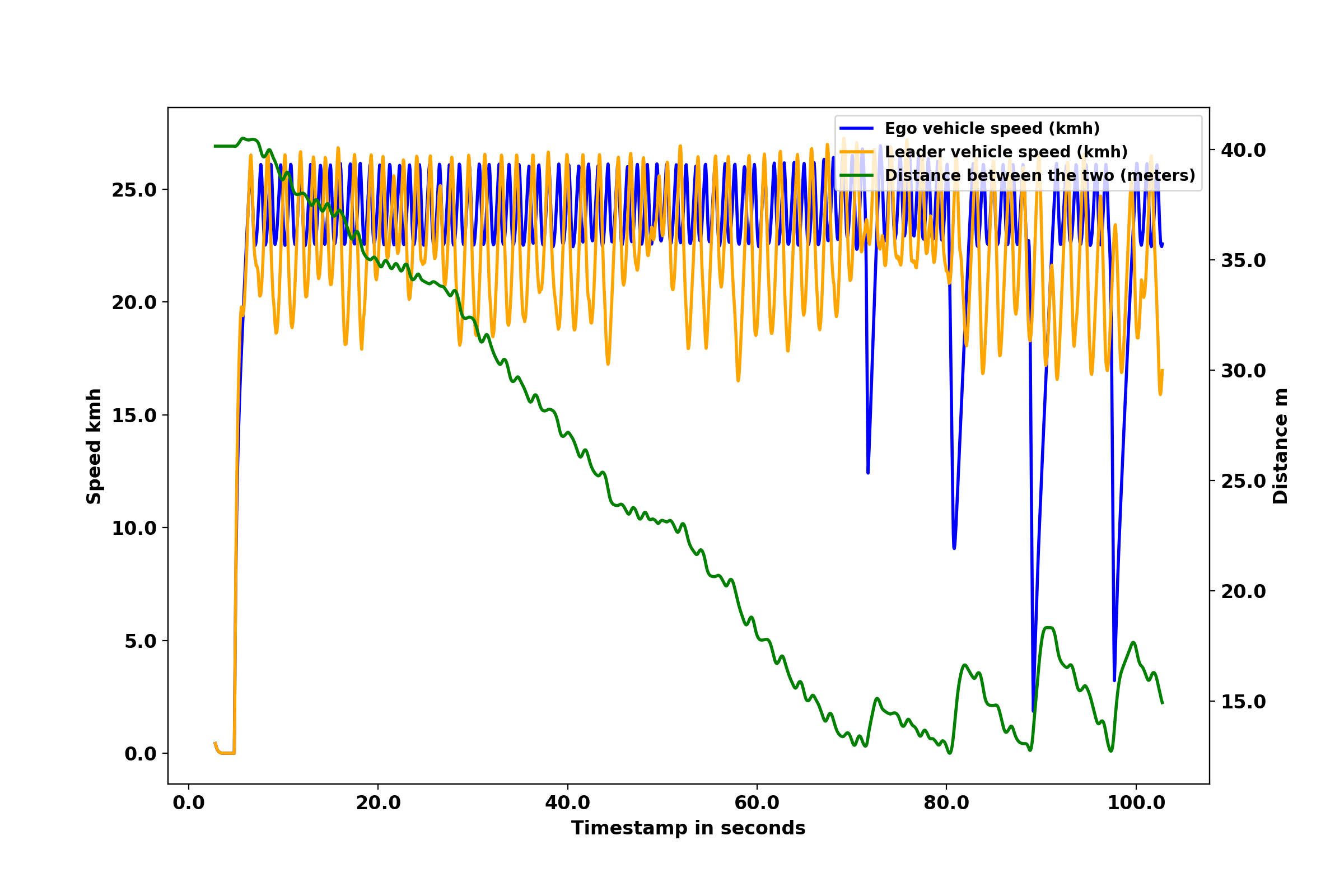}
    \centering
   \caption{The ego vehicle uses CAN log of moving vehicle while speed reading is being spoofing.}
    \label{fig:1st_speed}
    \end{subfigure}%
    \caption{Simulation results of the effectiveness of PID-based ACC for the scenario 1 - The ego vehicle drives at target speed of 25km/h and the lead vehicle drives at varying speed values. The figure shows that the distance between the two vehicles (in green) oscillates to keep a safe distance between the vehicles and avoid crashes. The used CAN Bus log is available in~\cite{9076852}.}
    
    \label{fig:Scenario1_output}
\end{figure*}

\begin{figure*}[tbp]
    \centering
     \begin{subfigure}[t]{0.33\textwidth}
        \centering
     \includegraphics[width=1.0\linewidth]{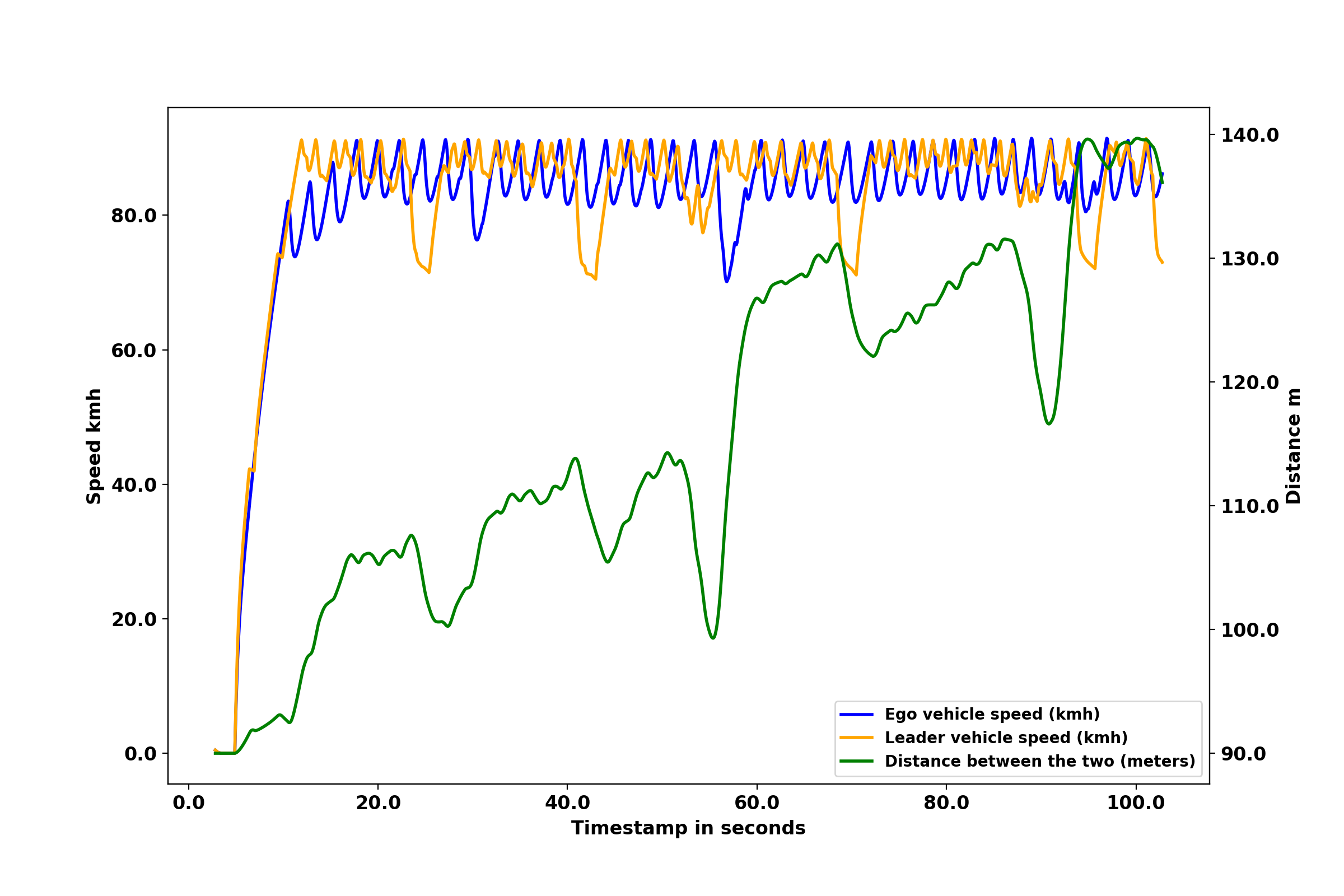}
    \caption{Simulation of normal driving of the ego and lead vehicle.}
    \label{subfig:scenario2normalriving}
    \end{subfigure}%
    ~
     \begin{subfigure}[t]{0.33\textwidth}
        \centering
     \includegraphics[width=1.0\linewidth]{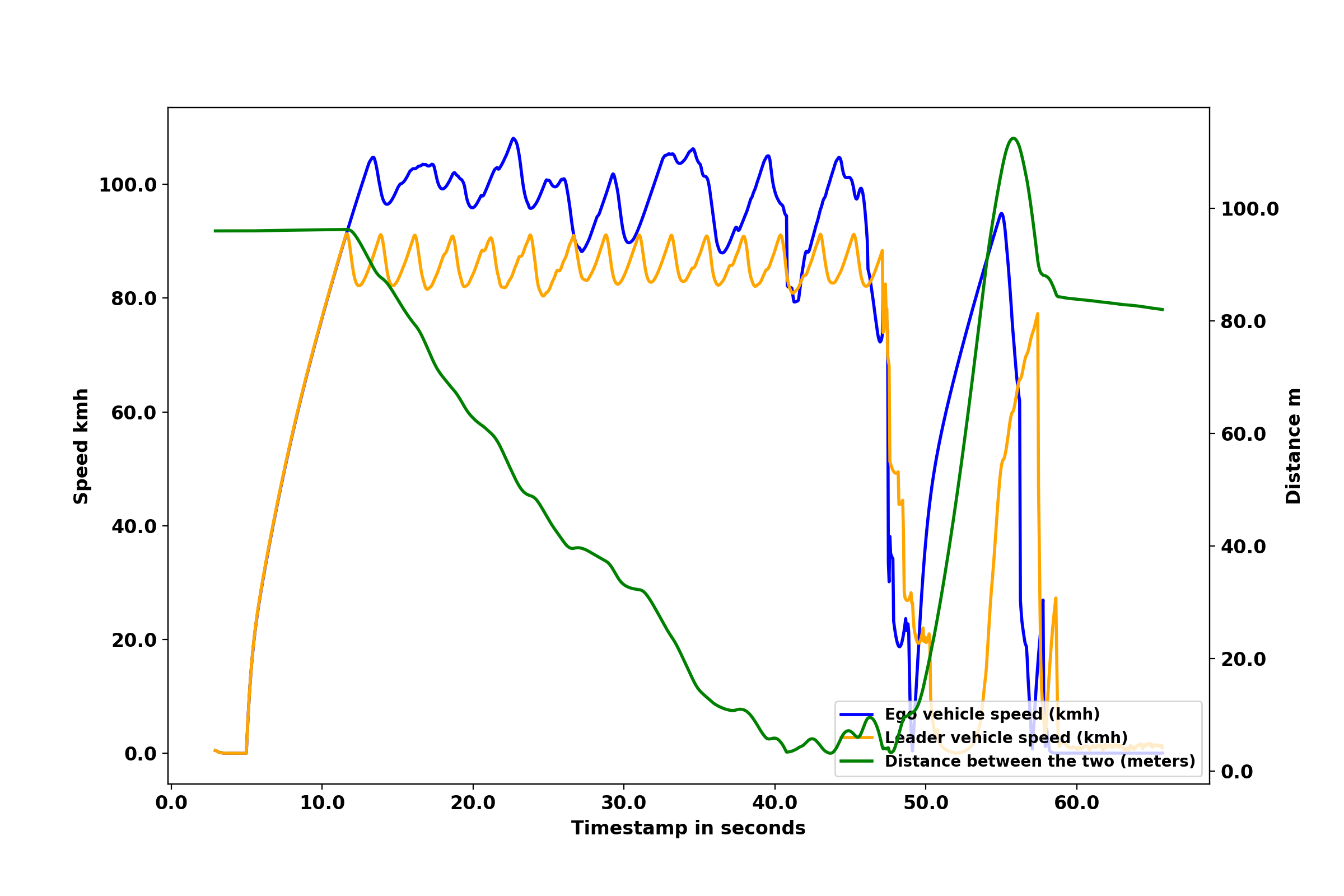}
    \caption{Simulation of normal driving of the lead vehicle and spoofing of the speed of the ego vehicle.}
    \label{subfig:scenario2speedspoofing}
    \end{subfigure}%
      ~
     \begin{subfigure}[t]{0.33\textwidth}
        \centering
     \includegraphics[width=1.0\linewidth]{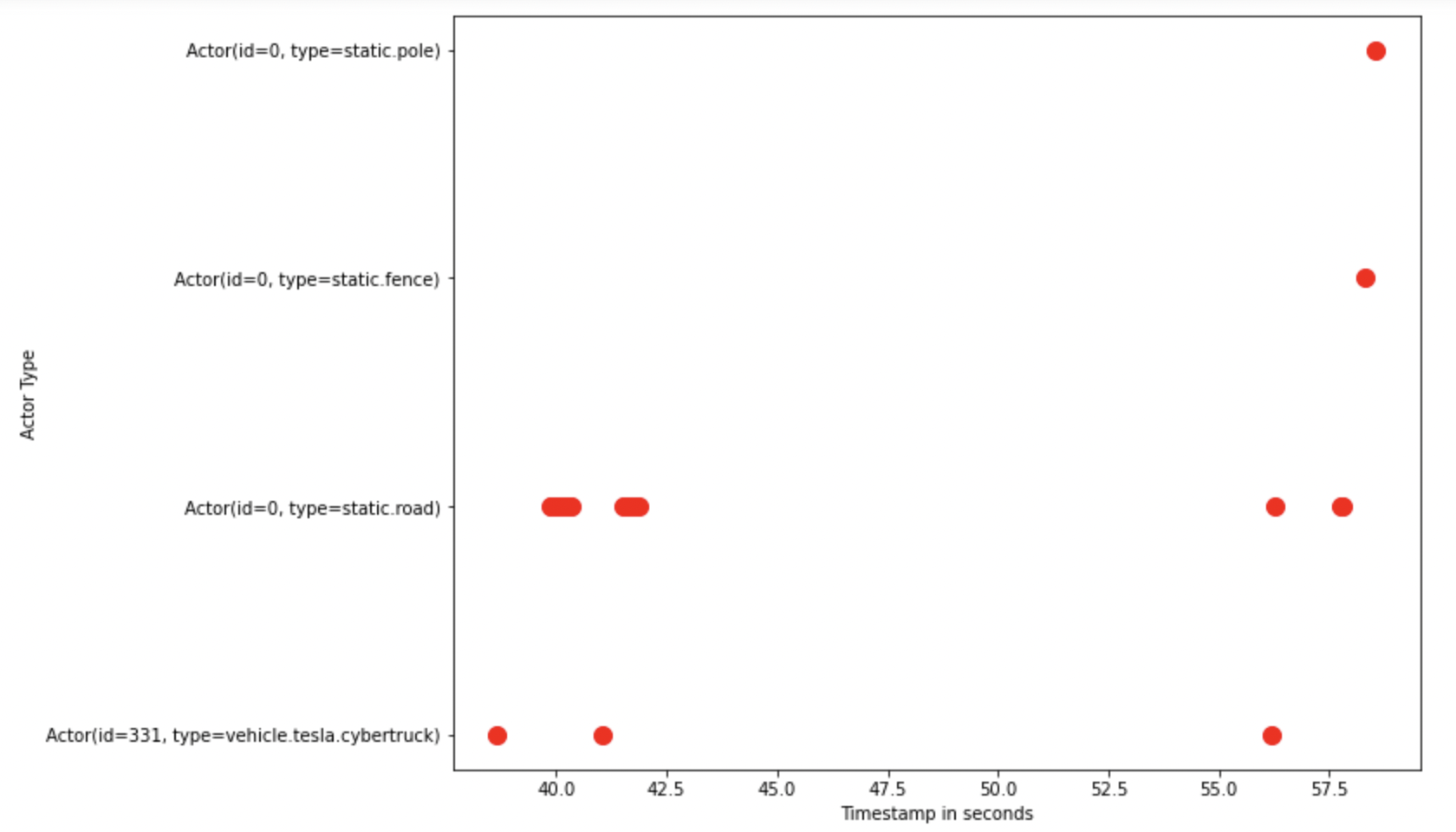}
   \caption{Collision frequency of the ego vehicle for the case of spoofing of the speed values.}
    \label{subfig:scenario2crashes}
    \end{subfigure}%
    \caption{Simulation results of the effectiveness of PID-based ACC for the scenario 2 - The ego and lead vehicles drive at target speed of 90km/h. Subfigure a shows that the distance between the two vehicles (in green) oscillates to keep a safe distance between the vehicles. Subfigure b shows that the distance between the two vehicles (in green) gets to low values between time steps 40 and 60. Subfigure c shows the crashes of the two vehicles. }
    \label{fig:Scenario2_output}
\end{figure*}

\begin{figure*}[tbp]
    \centering
     \begin{subfigure}[t]{0.31\textwidth}
         \centering
    \includegraphics[width=1.0\textwidth]{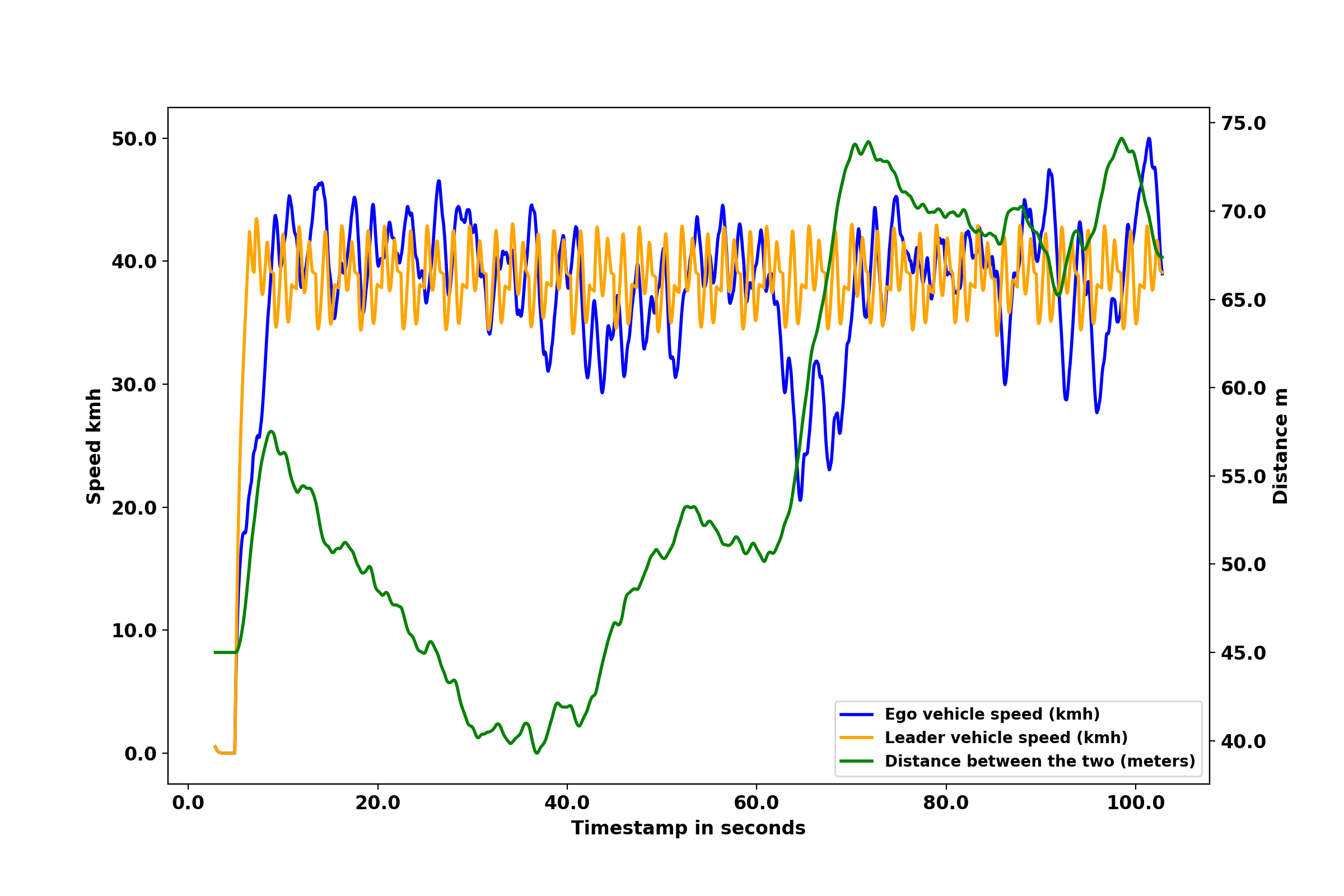}
    \caption{The ego vehicle speed is 40 km/h.}
    \label{fig:IDS_40}
\end{subfigure}
~
 \begin{subfigure}[t]{0.31\textwidth}
    \centering
    \includegraphics[width=1.0\textwidth]{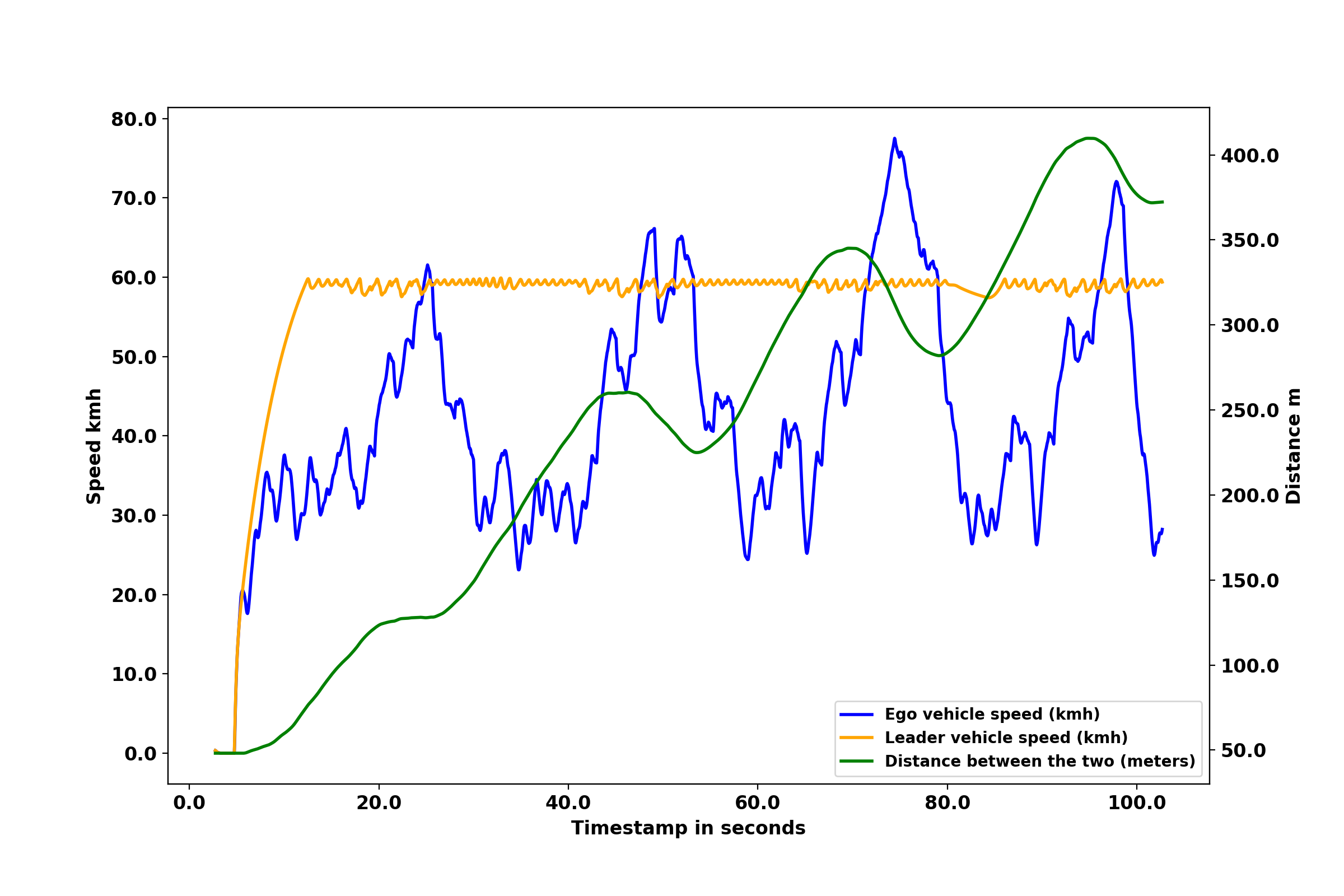}
    \caption{The ego vehicle speed is 60 km/h}
    \label{fig:IDS_60}
\end{subfigure}
~
 \begin{subfigure}[t]{0.31\textwidth}
    \centering
    \includegraphics[width=1.0\textwidth]{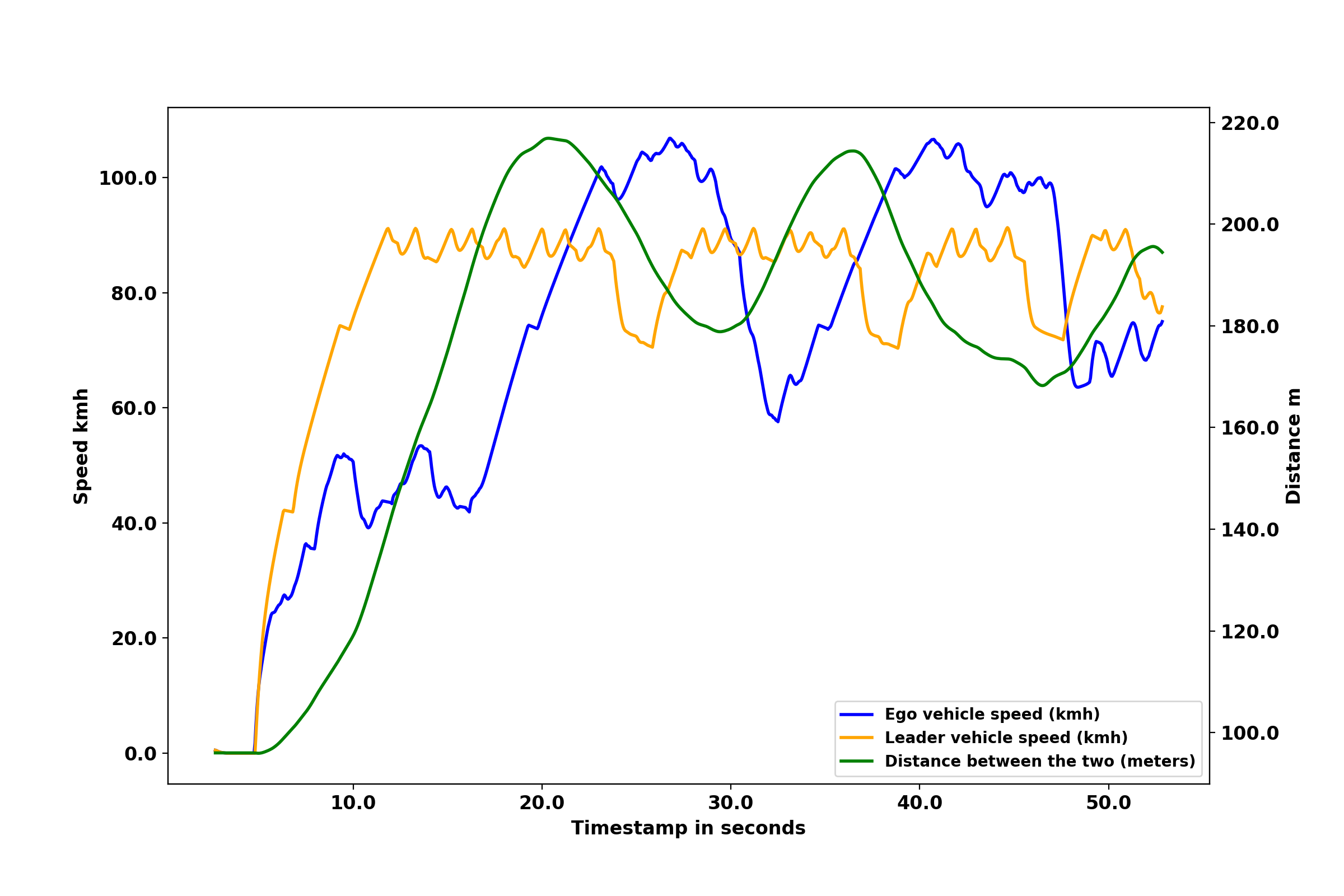}
    \caption{The ego vehicle speed is 90 km/h}
    \label{fig:IDS_90}
   \end{subfigure}
    \label{fig:Scenario3_output}
   \caption{Resilience response of ego vehicle while using the ACC-IDS. The figure shows that ACC-IDS reacts to messages injections correctly and avoids the crashes.}
\end{figure*}

\subsubsection{Results of simulating scenario 1}

In this scenario, we simulate an ego vehicle driving at target speed of 25km /h and a lead vehicle driving at speed of 0 km/h to 30 km/h and back to 0 km/h. We used a dataset~\cite{9076852} of a CAN Bus log of a moving car to mimic driving at speed of 0 km/h to 30 km/h and back to 0 km. 

Figure~\ref{fig:Scenario1_output} plots the speed of the ego vehicle, lead vehicle and the distance between the two vehicles as time progresses. The speed of the ego vehicle in subfigure~\ref{fig:1st_normal} is read from the log of a vehicle driving at speed increasing from 0 km/h to 30 km/h and back to 0 km/h again. The speed of the ego vehicle in subfigure~\ref{fig:1st_rpm} is read from the log of a vehicle driving at speed increasing from 0 km/h to 13 km/h and back to 0 km/h again, while the RPM reading is being spoofed. The speed of the ego vehicle in subfigure~\ref{fig:1st_speed} is read from the log of a vehicle driving at speed increasing from 0 km/h to 13 km/h and back to 0 km/h again, while the speed reading is being spoofed. 

The figure shows that the distance between the two vehicles (in green) oscillates to keep a safe distance between the vehicles. The ego vehicle accelerates and decelerates according to a command issued by the PID controller equation~\ref{PID_equation}. The distance between the two vehicles decreases multiple times but the ego vehicle applies emergency breaks or decreases the speed to avoid crashing into the lead vehicle. The spoofed RPM and speed datasets didn't cause the ego vehicle to crash into the lead vehicle. The figure shows that the \ac{PID}-based \ac{ACC} succeeds in this scenario to regulate the speed of the ego vehicle to avoid crashes.

\subsubsection{Results of simulating scenario 2}

In this scenario, we simulate an ego vehicle driving at target speed of 90 km/h and a lead vehicle driving at speed of 90 km/h. In addition, the speed sensor of the ego vehicle is spoofed to 10 km/h.  Figure~\ref{fig:Scenario2_output} plots the speed of the ego vehicle, lead vehicle and the distance between the two vehicles as time progresses. Subfigure a shows that the distance between the two vehicles (in green) oscillates to keep a safe distance between the vehicles. Subfigure b shows that the distance between the two vehicles (in green) gets to low values between time steps 40 and 60. Subfigure c shows the crashes of the two vehicles. The figure shows that \ac{PID}-based ACC succeed to regulate the speed of the ego vehicle in normal driving condition but fails to do so when the speed sensor is spoofed. The experiments was repeated for speed target of 40 km/h and 60 km/m with the same observation.\footnote{Interested reader may consult reference~\cite{mubarek2023}.}

\subsubsection{Results of simulating scenario 3} 

In this scenario, we simulate an ego vehicle driving at target speed of 90 km/h and a lead vehicle driving at speed of 90 km/h. In addition, the speed sensor of the ego vehicle is spoofed to 10 km/h and acts following a uniform distribution with probability of injection of spoofed message of 0.75. We changed the PID-based ACC controller to use a flag returned by the \ac{IDS}. The \ac{ACC} initiates cold break when an attack is returned by the IDS and applies the default behavior otherwise.   

Fig.~\ref{fig:IDS_90} shows the speed and distance of the ego vehicle and lead vehicle. The IDS was simulated with a success rate of 0.97\% and a response time of 149 ms as a single process with no threads. furthermore, Figure~\ref{fig:IDS_90} shows the distance between the ego and the lead vehicle increases drastically as the ego vehicle continues to apply the emergency brake to avoid crashing with the lead vehicle. 

\subsection{Impacts and limitations of the study}\label{ssec:impacts}

These simulation results demonstrate that extending ACC with ML-based real-time IDS would allow for mitigating cyber-attacks. The simulation covers three cases:
(1) vehicles drive at 40 km/h, and the spoofed speed is 5 km/h,
(2) vehicles drive at 60 km/h, and the spoofed speed is 10 km/h, and
(3) vehicles drive at 40 km/h, and the spoofed speed is 10 km/h.
The three scenarios show that spoofing the speed of the Ego vehicle misleads the ACC to accelerate, which causes accidents and that the ACC-IDS responds to the attacks and avoids accidents. 

The simulation of the three scenarios with the target speed of the ego vehicle and lead vehicle is set to 40 km/h~\cite{mubarek2023} shows that spoofing the speed of an ACC-equipped ego vehicle driving at a speed of 40km/h causes accidents when the spoofed value is 5 km/h and does not cause accidents when the value is 10km/h. We observe also that spoofing the speed of an ACC-equipped ego vehicle with a value of 10 km/h does not cause accidents when the speed of the vehicle is 40 km/h but causes an accident when the speed of the vehicle is 60 km/h or 90 km/h. Thus, the ACC-equipped vehicle fails to mitigate speed spoofing attacks based on the combination of the Ego vehicle's speed and the spoofed speed value.  

We used in the simulation study a spoofing detection success rate of 97\% and a response time of 1.026 seconds for the IDS. These parameters show that the proposed approach avoids accidents for the three simulated scenarios. An extensive simulation that varies the speed of the vehicle, the values of the spoofed speed, the IDS's success rate to detect injection, and the IDS's response time should be performed to set the boundaries of success of \ac{ACC}-{IDS} in mitigating message injection. 

\section{Conclusion}\label{sec:Conclusions}

Newer vehicles include ACCs that regulate the vehicle's speed for driver comfort and accident avoidance. Spoofing the vehicle speed communicated through the \ac{CAN} bus of the vehicle can mislead the \ac{ACC} to accelerate and crash the vehicle with the lead vehicle. The paper proposes extending the \ac{ACC} with a real-time \ac{IDS} capable of detecting speed spoofing attacks with reasonable response time, and detection rate. The CARLA simulation shows that the proposed extended ACC, called ACC-IDS, mitigates speed spoofing attacks as the ML-based \ac{IDS} triggers the brakes when an accident is imminent. The findings suggest exploring the integration of real-time \ac{IDS} into the resilience mechanisms to mitigate cyber-attacks on vehicles instead of using the estimation approach.

We will explore in the future the impact of the speed of the vehicle, the values of the spoofed speed, the IDS's success rate to detect injections, and the IDS's response time on the efficacy of the \ac{ACC}-{IDS} in mitigating cyber-attacks on the \ac{CAN} bus of connected vehicles.

\bibliographystyle{ieeetr}
\bibliography{ExtendACCwithIDS}

\end{document}